\documentclass{knawproc}

\def\etal{et~al.}
\def\spose#1{\hbox to 0pt{#1\hss}}
\def\lta{\mathrel{\spose{\lower 3pt\hbox{$\mathchar"218$}}
     \raise 2.0pt\hbox{$\mathchar"13C$}}}
\def\gta{\mathrel{\spose{\lower 3pt\hbox{$\mathchar"218$}}
     \raise 2.0pt\hbox{$\mathchar"13E$}}}
\def\Ha{H$\alpha$}

\usepackage{epsfig}
\begin{document}

\begin{opening}

\title{The nature of the 3CR radio galaxies at $z \sim 1$}

\author{Philip Best$^1$, Malcolm Longair$^2$ and Huub R\"ottgering$^1$}

\addresses{%
1. Sterrewacht, Huygens Lab, Postbus 9513, 2300\,RA Leiden, The
   Netherlands \\
2. Cavendish Labs, Madingley Road, Cambridge, CB3 0HE, United
   Kingdom 
}
 
\runningtitle{The 3CR galaxies at $z \sim 1$}
\runningauthor{Best et al.}

\end{opening}


\begin{abstract}

We present evidence that the 3CR radio galaxies at redshift $z \sim 1$ are
already very massive, highly dynamically evolved galaxies, which lie at
the heart of (proto--)cluster environments. Since nearby 3CR double radio
sources are generally found in more isolated surroundings, the galactic
environments of these galaxies must change dramatically with
redshift. Therefore, the original `uniform population, closed box'
interpretation of the infrared K--magnitude {\it vs} redshift relationship
no longer appears valid.  We propose a new interpretation: the powerful
radio galaxies selected at high and low redshift have different
evolutionary histories, but must contain a similar mass of stars, a few
times $10^{11} M_{\odot}$, and so conspire to produce the `passively
evolving' K$-z$ relation observed. We discuss this model in the context of
the current understanding of powerful radio sources and, in light of this
new model, we compare the K$-z$ relation of the 3CR galaxies with those
derived for lower power radio galaxies and for brightest cluster galaxies.
\end{abstract}


\section{Introduction}

\begin{figure}[!b]
\centerline{
\psfig{figure=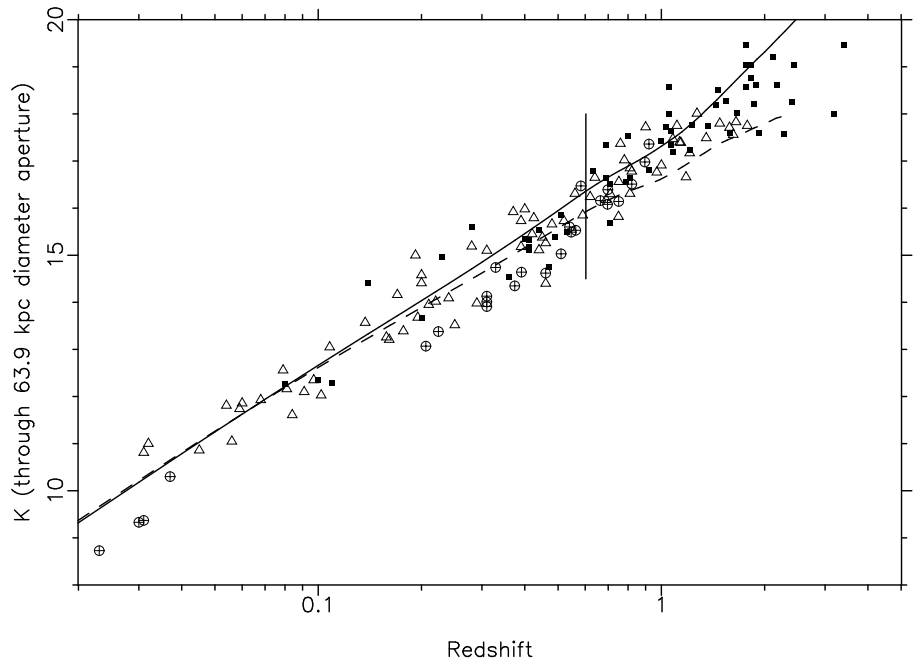,angle=-90,width=\textwidth,clip=}}
\caption{\label{kzdiag}The K$-z$ relation for 3CR radio galaxies (open
triangles; from Lilly and Longair 1984, Best \etal\ 1997a), 6C radio
galaxies (filled squares; from Eales \etal\ 1997), and brightest cluster
galaxies (crossed circles; from Arag\`on--Salamanca \etal\ 1993). The data
have been converted, where necessary, to a 63.9\,kpc metric aperture ($H_0
= 50$\,km\,s$^{-1}$\,Mpc$^{-1}$, $q_0 = 0.5$) using a radial light profile
appropriate for low redshift giant ellipticals (Lilly \etal\ 1984). The
solid and dashed lines show the relations expected for a non-evolving and
passively evolving ($z_{\rm f} = 5$) stellar populations respectively,
derived using the stellar synthesis codes of Bruzual and Charlot (1993)
and normalised to match the 3CR galaxies at low redshift. The vertical
line at redshift $z \sim 0.6$ provides a useful division between the high
and low redshift populations.}
\end{figure}

The revised 3CR sample is a complete sample of the brightest extragalactic
radio sources in the northern sky, selected at 178\,MHz (Laing \etal\
1983).  The host galaxies of the nearby `classical double', or FR\,II,
radio sources in the sample are giant elliptical galaxies. The infrared
K--magnitude {\it vs} redshift relation has been widely used as a tool for
investigating the evolution with cosmic epoch of the stellar populations
of luminous galaxies, since K--corrections, dust extinction corrections,
and the effect of any secondary star formation are all relatively
unimportant at near--infrared wavelengths. The remarkably small scatter
and slope of this relation for the 3CR radio galaxies is consistent with
passive, `closed--box' evolution of a single population of giant
elliptical galaxies which formed at large redshift (see
Figure~\ref{kzdiag}), suggesting that the distant 3CR galaxies are
precisely the same objects as nearby 3CR galaxies, but observed at an
epoch when their stellar populations were younger (Lilly and Longair
1984).

\nocite{bes97d}

We have selected an almost complete sample of 28 radio galaxies from the
3CR sample, with redshifts in the range $0.6 < z < 1.8$. These galaxies
were observed for one orbit each in each of two colours (at rest--frame
wavelengths of approximately 315 and 475\,nm) using the HST, and for about
an hour in the near--infrared J and K--bands using UKIRT, with 1 arcsecond
angular resolution. A detailed description of the observations, and the
reduced images, can be found in Best \etal\ (1997a). 

\section{The host galaxies of 3CR radio sources}

\subsection{The origin of the near--infrared emission}

In 1987, McCarthy \etal\ and Chambers \etal\ discovered that the optical
(rest--frame ultraviolet) emission of powerful distant radio galaxies is
elongated and aligned along the direction defined by the radio lobes. The
high resolution HST images of the 3CR galaxies show this clearly, with
some galaxies having many bright optical knots strung along the radio jet
axis (Best \etal\ 1997a). A number of models have been proposed for this
`alignment effect', the three most promising being jet--induced star
formation (e.g. Rees 1989), scattering of light from an obscured quasar by
electrons or dust (Cimatti \etal\ 1996 and references therein), and
nebular continuum emission (Dickson \etal\ 1995). Some combination of all
of these three processes is most likely to be occurring. A complete
discussion of these issues is not possible in the space available here,
and can be found in many other contributions to this volume. It is
important to note, however, that each of these alignment mechanisms
produces relatively flat spectrum emission, and so in the K--band this
aligned emission will be dominated by that of the old stellar
population. Best \etal\ (1997b) showed that, on average, only about 10\%
of the K--band flux density of the 3CR radio galaxies at redshift one is
associated with the aligned component.

\begin{figure}
\centerline{
\psfig{figure=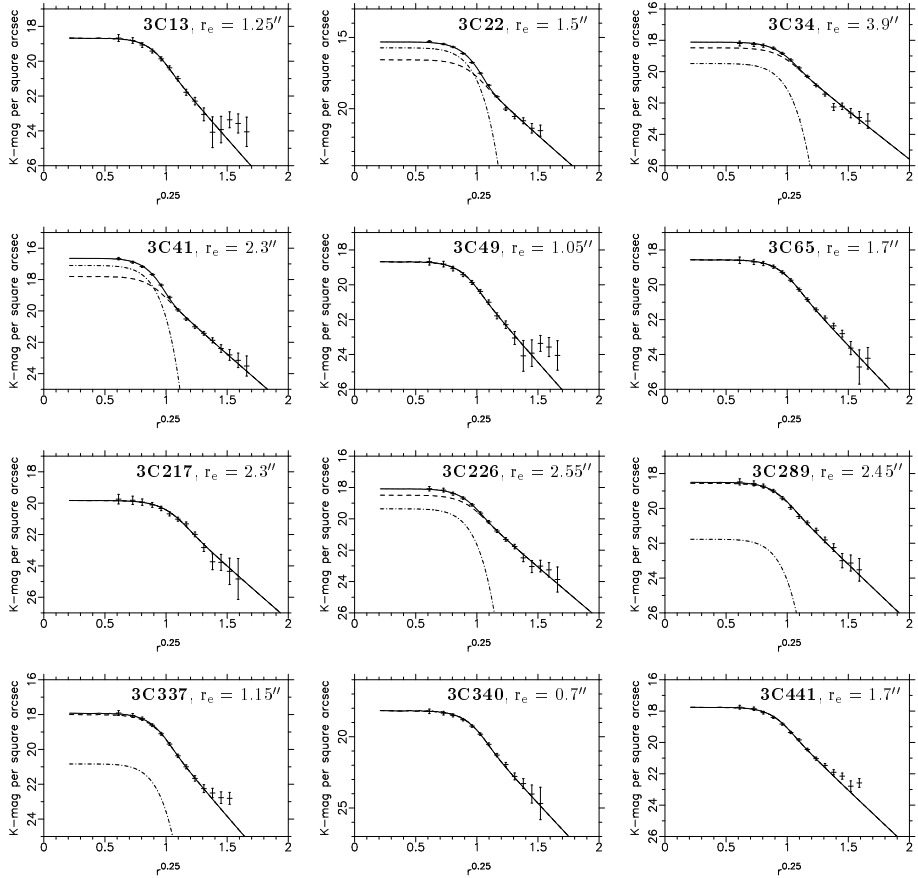,width=\textwidth,clip=}}
\caption{\label{vauc} Fits to the K--band radial intensity profiles of 12
3CR galaxies from the sample, using the sum of an unresolved point source
(dash--dot line) and a de Vaucouleurs profile (dashed line). For each of
the profiles, the effect of seeing has been taken into account. The sum of
the two components is indicated by the solid line. In many cases the best
fit does not involve a point source component, and so only the solid line
is shown.}
\end{figure}

An investigation of the infrared radial intensity profiles of these 3CR
galaxies shows that, in general, they can be well--matched using de
Vaucouleurs' law, with little requirement for any contribution from
unresolved nuclear emission. A detailed description of the fitting
procedure can be found in Best \etal\ (1997b); in Figure~\ref{vauc}, we
show examples of the fits for 12 of the galaxies. There is significant
evidence for a contribution to the K--band flux density from an unresolved
nuclear emission source in only two of the sources in the sample, 3C22 and
3C41, providing about 50\% and 30\% respectively of the K--band light in
these cases. Interestingly, Rawlings \etal\ (1995) have already proposed
that 3C22 contains a reddened quasar, based upon the red colour of its
spectral energy distribution and the detection of broad \Ha\ emission. For
the remaining sources in the sample, the best fit to the K--band radial
profile has $\lta 10\%$ of the total flux density associated with a point
source component, and in each case this is consistent, within the 90\%
confidence limits, with being zero.

A third possible AGN--related contribution to the emission of these
galaxies is line emission, but for the redshifts of the present sample
there are no major emission lines in the K--band (\Ha\ enters the passband
beyond $z \approx 1.9$), and so line emission contributes typically $\lta
5\%$ of the K--band flux density. We conclude that the K--band flux
density of the 3CR radio galaxies at redshift $z \sim 1$ is dominated by
emission from their old stellar populations.

\subsection{Fundamental parameters of the 3CR radio galaxies} 

\begin{figure}
\centerline{
\psfig{figure=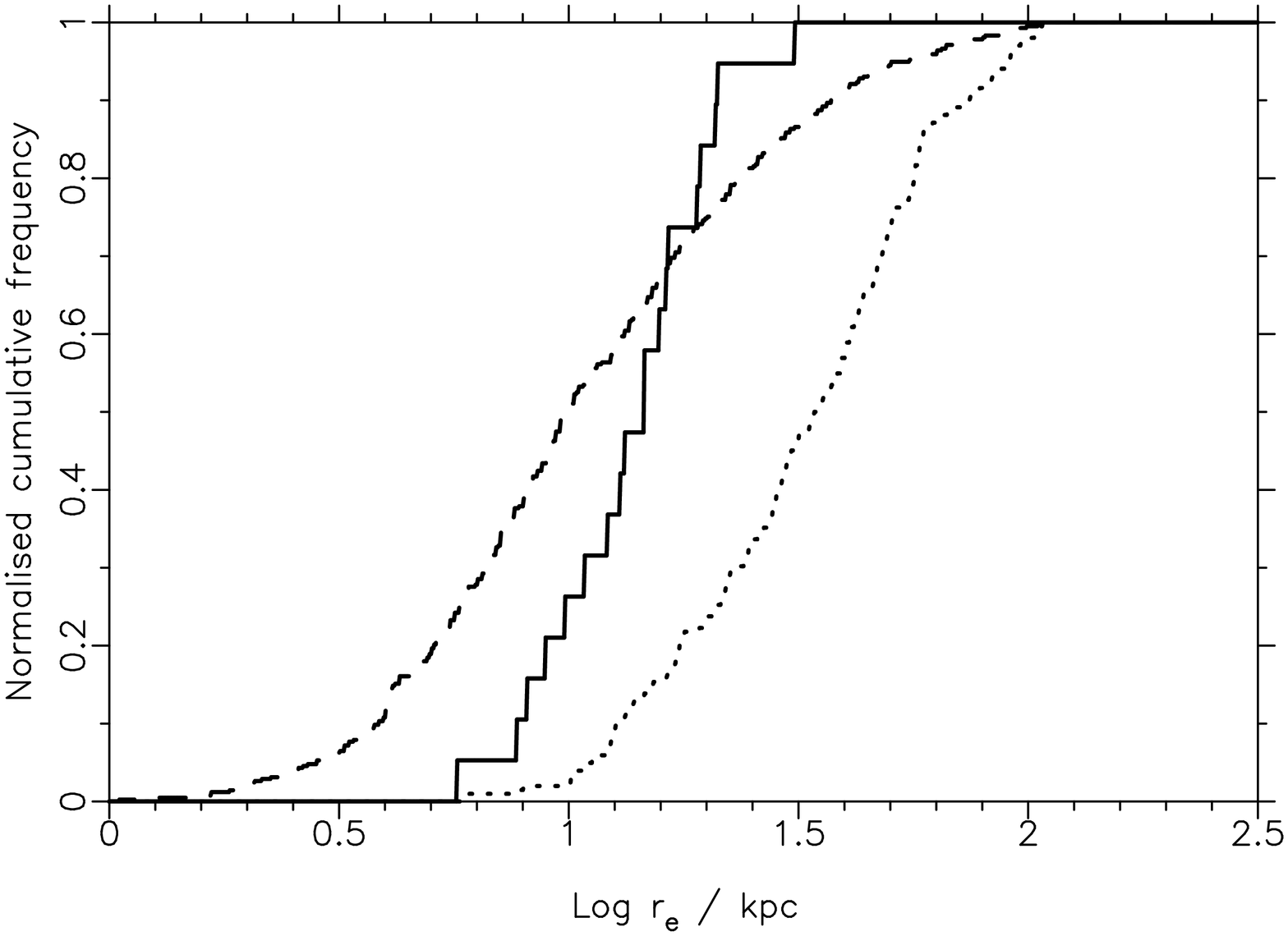,clip=,angle=-90,width=7.5cm}
}
\caption{\label{replot} Normalised cumulative frequency distributions
of $r_{\rm e}$ for high redshift 3CR galaxies (solid line), low redshift
brightest cluster galaxies (dotted line) and low redshift elliptical
galaxies, excluding brightest cluster galaxies (dashed line). The latter
two samples are taken from Schombert (1987).}
\end{figure}

The characteristic radii, $r_{\rm e}$, of the distant 3CR galaxies, as
determined from the de Vaucouleurs fits, are generally large with a mean
value of $14.6 \pm 1.4$\,kpc ($q_{\rm 0} = 0.5$). In terms of standard
cannibalism models (e.g. Hausman and Ostriker 1978), the characteristic
radius of a galaxy can be used as a measure of its dynamical evolutionary
history, the galaxies with the largest characteristic radii having
undergone the most mergers. The mean value obtained for the distant 3CR
galaxies is significantly larger than that of the low redshift elliptical
galaxies from the sample of Schombert (1987), which have $\overline{r_{\rm
e}} = 8.2 \pm 1.0$\,kpc, and is only about a factor of two smaller than
the mean value for the low redshift brightest Abell cluster galaxies in
the same sample ($\overline{r_{\rm e}} = 32.7 \pm 1.1$\,kpc; see
Figure~\ref{replot}). Kormendy (1977) showed that the integrated luminosity
($\cong$ mass) of bright elliptical galaxies increases approximately as
$r_{\rm e}^{0.7}$, and so the distant 3CR radio galaxies are only about a
factor of two less massive than present day brightest cluster galaxies
(BCGs). These galaxies must be very massive and highly dynamically
evolved, even by a redshift of one.

\begin{figure}
\centerline{
\psfig{figure=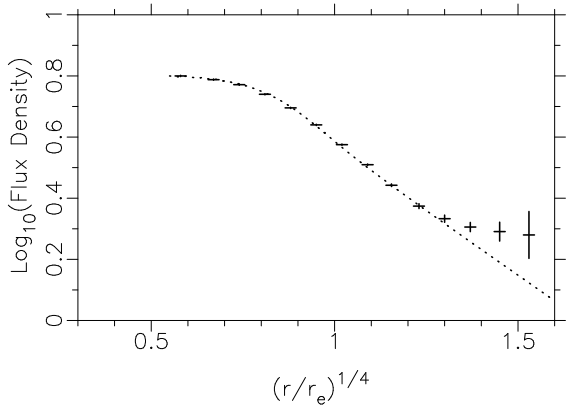,angle=-90,clip=,width=7.5cm} 
}
\caption{\label{halo} A combined K--band radial intensity profile for the
12 3CR galaxies with redshifts $0.6 < z < 1.8$ and characteristic radii
$1.0 \le r_{\rm e} \le 2.0$.}
\end{figure}

Interestingly, at large radii a number of the 3CR galaxies show an excess
of emission over that expected from the de Vaucouleurs profile. To
increase the signal--to--noise of this feature, the galaxies with $1.0'' <
r_{\rm e} < 2.0''$ were scaled by their characteristic radii and their
profiles summed. Galaxies with smaller characteristic radii were not
included because the effects of the seeing profile may still be important
at a radius of about 3\,$r_{\rm e}$, whilst larger galaxies were discarded
because of low signal--to--noise in their outer regions. The combined
profile is shown in Figure~\ref{halo}, and clearly shows an excess of
emission beyond a radius corresponding to $r \sim 35$\,kpc. This is the
same radius as that beyond which halos are seen around cD galaxies in
nearby clusters (e.g. Oemler 1976). Two observations dictate that this
excess emission is not associated with the alignment effect: (i) well over
95\% of the aligned emission of these galaxies lies within a radius of
35\,kpc on the HST images; (ii) if the sample is split into galaxies with
a strong and with a weak aligned component, the halo is seen with equal
strength in each.

\section{The cluster environments of the 3CR galaxies at redshift $z \sim 1$}

We have presented evidence that the distant 3CR galaxies are giant
elliptical galaxies which: (i) have absolute K--magnitudes comparable to
BCGs at redshifts $z \gta 0.7$; (ii) have large characteristic radii, and
are highly dynamically evolved; (iii) contain old stellar populations;
(iv) show, in at least some cases, an excess of emission at large radii
similar to the halos seen around cD galaxies.

Further evidence also suggests that powerful radio galaxies at high
redshift ($z \gta 0.5$) tend to lie in young or forming cluster
environments (see Best \etal\ 1997b for a review).  Specifically: (i)
measures of the galaxy cross--correlation function (Yates \etal\ 1989) and
an Abell clustering classification (Hill and Lilly 1991) both indicate
that powerful FR\,II sources at intermediate redshift ($z \sim 0.5$)
belong to environments of Abell class 0 richness or greater; (ii) powerful
extended X--ray emission from the vicinity of a number of high redshift
3CR radio sources has been associated with cooling flows in relatively
dense intracluster media (e.g. Crawford and Fabian 1996); (iii) [OII]~3727
emission line imaging has revealed companion galaxies near many of the
radio sources (McCarthy 1988); (iv) multicolour imaging of the fields of
powerful radio sources reveals a significant excess of galaxies with the
red colours expected of a passively evolving population coeval with the
radio source (e.g. Dickinson 1997); (v) the large Faraday depolarisation
and rotation measures of these radio sources require a dense, ionised
surrounding medium (e.g. Carilli \etal\ 1997); (vi) spectroscopic studies
of the fields of some distant 3CR galaxies show them to lie in at least
moderately rich clusters (Dickinson 1997).

In contrast, the amplitudes of the spatial cross--correlation functions
for galaxies in the vicinity of nearby 3CR FR\,II radio sources are
similar to those for normal elliptical galaxies (Prestage and Peacock
1988), whilst the host radio galaxies themselves have optical luminosities
and characteristic sizes significantly less than those of first ranked
Abell cluster galaxies (Lilly and Prestage 1987). Thus, powerful FR\,II
radio galaxies at low redshift tend to lie in isolated environments or in
small groups, whilst the distant 3CR radio galaxies ($z \gta 1$) appear to
be associated with the brightest galaxies in rich cluster or
proto--cluster environments. This dramatic change in the galactic
environments of the 3CR radio galaxies, from richer to poorer environments
with increasing cosmic time, indicates that the `uniform population'
interpretation of the K$-z$ relationship, in which the host galaxies of
distant 3CR radio galaxies evolve passively to become the equivalent of
the hosts of nearby 3CR galaxies, cannot be correct.

\section{The nature of powerful radio sources}

To produce a powerful radio source, three essential ingredients are
required: (i) a supermassive central black hole; (ii) a plentiful supply
of gas to fuel the black hole; (iii) a dense surrounding medium to
minimise adiabatic expansion of the radio lobes. The 3CR radio galaxies at
$z \sim 1$ release huge fluxes of kinetic energy in the form of
relativistic jets, corresponding to the Eddington limiting luminosity of
black holes with $M \sim 10^8 -10^9\,M_{\odot}$ (Rawlings and Saunders
1991).  Theoreticians have long argued that the masses of the central
black holes should be roughly proportional to the masses of the host
galaxies (e.g. Soltan 1982; Efstathiou and Rees 1988), and some evidence
for such a correlation has recently been found for massive black holes in
the nuclei of nearby galaxies (e.g. Kormendy and Richstone 1995). Massive
central cluster galaxies would therefore be expected to host the most
powerful central engines. In high redshift proto--cluster or young cluster
environments, there will also be a plentiful supply of disturbed
intracluster gas to fuel the central engine and confine the radio
lobes. Furthermore, Ellingson \etal\ (1991) found that the velocity
dispersions of galaxies around powerful distant radio sources are
significantly lower (400 to 500\,km\,s$^{-1}$) than those in comparably
rich low redshift clusters (500 to 1000 km\,s$^{-1}$). These smaller
relative velocities increase the efficiency of galaxy mergers, which may
trigger the onset of the radio source (e.g. Heckman \etal\ 1986). Thus,
massive galaxies in young cluster environments at high redshift possess
all of the necessary ingredients for producing the most powerful radio
sources.

Why then are the most powerful FR\,II radio sources at the present epoch
not also associated with central cluster galaxies?  The radio sources
hosted by nearby BCGs are almost invariably `edge--darkened' FR\,I type
radio sources; their radio jets are of much lower kinetic power than the
FR\,IIs and, according to current ideas, are strongly decelerated in the
inner kiloparsec by entrainment of surrounding material (Laing 1993 and
references therein), producing lower luminosity radio sources.  Since
neither the black hole mass nor the density of the cluster environment can
decrease with cosmic epoch, the availability of fuelling gas must be the
factor which limits the jet--powers of these sources.

Rich regular clusters at low redshift have had time to virialise the
spatial distribution of the galaxies, and for the intracluster gas to take
up an equilibrium configuration within the cluster gravitational
potential. The high velocity dispersion of the galaxies in these clusters
greatly reduces the merger efficiency. Furthermore, studies of the
Butcher--Oemler effect indicate that, compared to high redshift clusters,
there is a dearth of gas--rich galaxies close to the centre of low
redshift clusters which might merge with, and fuel, the central galaxy.
The evolution and virialisation of low redshift clusters is therefore
likely to result in a more restricted fuelling supply for the central
engines of BCGs. Instead, the most powerful FR\,II radio sources at small
redshift are generally found in small groups of galaxies: in these groups,
velocity dispersions are small, mergers relatively frequent, and gas--rich
galaxies plentiful. Occasionally, dramatic mergers will occur in low
redshift BCGs, and we ascribe exceptional sources such as Cygnus A
(e.g. Fosbury, this volume) to this process.

These considerations suggest an explanation for the constancy of the
stellar masses of the 3CR galaxies throughout the redshift range $0.03 \le
z \le 1.8$.  This mass may be interpreted as the typical mass which a
giant elliptical galaxy attains before its galactic and gaseous
environment is virialised.  Galaxies of mass greater than a few times
$10^{11} M_{\odot}$ will have undergone sufficient dynamical evolution
that they generally live in virialised environments where the reduced
supply of fuelling gas to the central regions results in FR\,I sources
being formed.  This leads naturally to a correlation between the redshift
of a 3CR FR\,II galaxy and the richness of its environment: galaxies in
highly overdense environments accumulate matter the fastest, and therefore
reach this `FR\,II upper mass limit' at early cosmic epochs, whilst those
which lie in less dense environments evolve more slowly and can form
powerful FR\,II radio galaxies at the current epoch.

\section{Cosmic conspiracies in the K$-z$ relations}

In light of these ideas, the K$-z$ relation of the 3CR galaxies can be
compared with those derived for other samples of galaxies, such as the 6C
radio galaxies, which are about a factor of five less powerful radio
sources than the 3CR galaxies at a given redshift (Eales and Rawlings
1996; Eales \etal\ 1997) and a sample of BCGs (Arag\'on--Salamanca \etal\
1993). In these cases, their mean K$-z$ relations are broadly consistent
with non--evolving stellar populations (see Figure~\ref{kzdiag}). This is
remarkable, since undoubtedly their stellar populations must have evolved
significantly between redshift $z \sim 1$ and the present epoch. Indeed,
the V$-$K colours of these galaxies are bluer at $z \gta 0.5$ than they
are nearby, consistent with passively evolving stellar populations
(Arag\'on--Salamanca \etal\ 1993).

At high redshift ($z \sim 1$), the 6C radio galaxies have absolute
K--magnitudes which are about 0.6 magnitudes fainter than those of the 3CR
sample (Eales and Rawlings 1996; Eales \etal\ 1997; however, see also
McCarthy, this volume) whilst at low redshift the two samples are equally
luminous in the K--waveband. Eales \etal\ suggested that this effect might
be associated with emission of the active galactic nucleus contributing
directly or indirectly to the K--band flux density of the distant 3CR
sources. The analysis of our HST and UKIRT observations presented in
Section~2 demonstrates that these mechanisms can account for at most $\sim
0.3$ magnitudes of the difference.  The greater K--band luminosities of
the 3CR galaxies as compared with the 6C galaxies at high redshift must
indicate, therefore, that they contain a greater mass of stars. It follows
that, if there is a correlation between stellar mass and central black
hole mass, the 6C galaxies possess less powerful central engines than the
3CR galaxies at high redshift, thus accounting for their lower radio
luminosities. The overlap between the two samples on the K$-z$ relation
would then arise from the scatter in the correlation between the stellar
mass of the galaxy and the mass of the black hole (e.g. Kormendy and
Richstone 1995).

At redshifts $z \lta 0.6$, the beam powers of all radio galaxies are well
below the Eddington limit (Rawlings and Saunders 1991) indicating that
their radio luminosities are determined predominantly by the availability
of fuelling gas in the host galaxy and its environment, rather than by the
black hole mass. A weaker correlation would therefore be expected between
galaxy mass and radio luminosity at these redshifts, accounting for the
similarity of the K--magnitudes for the low redshift 3CR and 6C galaxies.

Although the 3CR galaxies and BCGs are equally bright at redshifts $z \sim
1$, at low redshifts BCGs are up to a magnitude brighter in absolute
K--magnitude than the 3CR galaxies.  Since the stellar populations of the
BCGs must have evolved over this redshift interval in a similar way to
those of the 3CR galaxies, the shape of their K$-z$ relation must reflect
the fact that BCGs continue to accumulate matter through mergers with
massive cluster galaxies, and gas infall.  Hierarchical clustering models
for structure formation have suggested that the mass of BCGs increases by
a factor of 3--4 between a redshift of one and the present epoch
(Kauffmann 1995, Arag\'on--Salamanca \etal\ 1997).

For the BCGs, and also for the 6C galaxies, this growth of the mass of the
galaxies between $z \sim 1$ and $z = 0$ increases their absolute
magnitudes, thereby compensating for the dimming of their stellar
populations --- the two effects conspire to give rise to apparently simple
`no evolution' tracks. The 3CR galaxies at high redshift also lie at the
centre of galaxy clusters and so their masses would be expected to
increase with cosmic epoch just like the BCGs; the `passively evolving'
K$-z$ relation of these galaxies suggests, however, that this does not
occur. This, too, is a conspiracy.  The galaxies sampled at high and low
redshifts do not form a uniform population, as is indicated by the
dramatic change in their galactic environments with redshift. The apparent
passive evolution disguises an important result: the most powerful FR\,II
radio sources at all redshifts $z \lta 1.5$ contain approximately the same
mass of stars, a few times $10^{11} M_{\odot}$.  

\section{Conclusions}

We have presented evidence that powerful distant radio sources lie in
(proto) cluster environments, and so provide a unique tracer of the most
massive systems in the difficult redshift interval $1 \lta z \lta 3$.
Studies of the evolutionary state of such massive collapsed (or
collapsing) structures at early epochs will provide important constraints
upon theories of structure formation and galaxy evolution.

We have also demonstrated that the commonly--used `closed--box, uniform
population' galaxy evolution models are not appropriate for
interpretations of the K$-z$ relations.  Instead, the shapes of these
relations can be used to provide information about the merger histories of
massive galaxies in clusters.
\bigskip

\begin{acknow}
This work was supported in part by the Formation and Evolution of Galaxies
network set up by the European Commission within its TMR programme, and by
a programme subsidy granted by the Netherlands Organisation for Scientific
Research (NWO).
\end{acknow}


\begin{references}
\bibitem[]{ara93} {Arag{\'o}n--Salamanca}~A., Ellis~R., Couch~W.,
Carter~D., 1993, MNRAS, 262, 764

\bibitem[]{ara97} {Arag{\'o}n--Salamanca}~A., Baugh~C.M., Kauffmann~G.,
MNRAS, submitted.

\bibitem[]{bes97c} Best~P.N.,  Longair~M.S.,    R{\"o}ttgering~H.J.A.,
1997a, MNRAS, in press, astro-ph/9707337

\bibitem[]{bes97d} Best~P.N.,  Longair~M.S.,    R{\"o}ttgering~H.J.A.,
1997b, MNRAS, in press, astro-ph/9709195

\bibitem[]{bru93} Bruzual~G., Charlot~S., 1993, ApJ, 405, 538
\bibitem[]{car97} Carilli~C.L.,  R{\"o}ttgering~H.J.A.,  {van Ojik}~R.,
Miley~G.K.,    {van Breugel}~W. J.M., 1997, ApJ Supp., 109, 1

\bibitem[]{cha87} Chambers~K.C.,  Miley~G.K.,    {van Breugel}~W. J.M.,
1987, Nat, 329, 604 

\bibitem[]{cim96} Cimatti~A., Dey~A., {van Breugel}~W., Antonucci~R.,
Spinrad~H., 1996, ApJ, 465, 145

\bibitem[]{cra96b} Crawford~C.S.,  Fabian~A.C.,  1996, MNRAS, 282, 1483

\bibitem[]{dic97a} Dickinson~M.,  1997, in Tanvir~N.R.,
Arag{\'o}n-Salamanca~A.,   Wall~J.V.,  eds, HST and the high redshift
Universe. Singapore: World Scientific

\bibitem[]{dic95} Dickson~R., Tadhunter~C., Shaw~M., Clark~N.,
Morganti~R., 1995, MNRAS, 273, L29

\bibitem[]{eal96} Eales~S.A., Rawlings~S.,  1996, ApJ, 460, 68

\bibitem[]{eal97} Eales~S.A.,  Rawlings~S.,  {Law--Green}~D.,  Cotter~G.,
Lacy~M.,  1997, MNRAS, in press, astro-ph/9701023

\bibitem[]{efs88} Efstathiou~G.,  Rees~M.,  1988, MNRAS, 230, 5P

\bibitem[]{ell91} Ellingson~E.,  Green~R.F.,    Yee~H. K.C.,  1991, ApJ,
378, 476 

\bibitem[]{hau78} Hausman~M.A., Ostriker~J.P., 1978, ApJ, 224, 320

\bibitem[]{hec86} Heckman~T.M.,  Smith~E.P.,  Baum~S.A.,  
{van Breugel}~W.J.M., Miley~G.K.,  Illingworth~G.D.,  Bothun~G.D.,    
Balick~B.,  1986, ApJ, 311, 526

\bibitem[]{hil91} Hill~G.J.,  Lilly~S.J.,  1991, ApJ, 367, 1

\bibitem[]{kau95a} Kauffmann~G.,  1995, MNRAS, 274, 153

\bibitem[]{kor77} Kormendy~J.,  1977, ApJ, 217, 406

\bibitem[]{kor95} Kormendy~J.,  Richstone~D.,  1995, ARA\&A, 33, 581

\bibitem[]{lai93} Laing~R.A., 1993, in Burgarella~D., Livio~M., 
{O'Dea}~C.P.,  eds, Astrophysical Jets. Cambridge University Press, 
Cambridge, p.95

\bibitem[]{lai83} Laing~R.A., Riley~J.M., Longair~M.S., 1983, MNRAS, 
204, 151

\bibitem[]{lil84a} Lilly~S.J.,  Longair~M.S.,  1984, MNRAS, 211, 833

\bibitem[]{lil84c} Lilly~S.J., McLean~I.S., Longair~M.S., 1984, MNRAS,
209, 401

\bibitem[]{lil87} Lilly~S.J.,  Prestage~R.M.,  1987, MNRAS, 225, 531

\bibitem[]{mcc88} McCarthy~P.,  1988, Ph.D. thesis, University of 
California, Berkeley

\bibitem[]{mcc87} McCarthy~P., {van Breugel}~W., Spinrad~H.,  
Djorgovski~S., 1987, ApJ, 321, L29

\bibitem[]{oem76} Oemler~A.,  1976, ApJ, 209, 693

\bibitem[]{pre88} Prestage~R.M., Peacock~J.A., 1988, MNRAS, 230, 131

\bibitem[]{raw95} Rawlings~S., Lacy~M., Sivia~D.S., Eales~S.A., 1995,
MNRAS, 274, 428

\bibitem[]{raw91b} Rawlings~S.,  Saunders~R.,  1991, Nat, 349, 138

\bibitem[]{ree89} Rees~M.J., 1989, MNRAS, 239, 1P

\bibitem[]{sch87} Schombert~J.M.,  1987, ApJ Supp., 64, 643

\bibitem[]{sol82} Soltan~A.,  1982, MNRAS, 200, 115

\bibitem[]{yat89} Yates~M.G., Miller~L., Peacock~J.A., 1989, MNRAS, 240, 129
\end{references}
\end{document}